\documentclass[aps,twocolumn,groupedaddress,showpacs]{revtex4}
\usepackage{graphicx}

\begin{document}

\title[]
{Tricritical point of {\boldmath{$J_1^{~}$-$J_2^{~}$}} Ising model on hyperbolic lattice}

\author{R.~Krcmar$^1$, T.~Iharagi$^2$, A.~Gendiar$^{1,3}$, and T.~Nishino$^2$}
\affiliation{$^1$Institute of Electrical Engineering, Centre of Excellence CENG,
Slovak Academy of Sciences, D\'{u}bravsk\'{a} cesta 9, SK-841~04, Bratislava, Slovakia\\
$^2$Department of Physics, Graduate School of Science, Kobe University,
Kobe 657-8501, Japan\\
$^3$Institute for Theoretical Physics C, RWTH Aachen University, D-52056 Aachen, Germany}

\date{\today}

\begin{abstract}
A ferromagnetic-paramagnetic phase transition of the two-dimensional frustrated
Ising model on a hyperbolic lattice is investigated by use of the corner
transfer matrix renormalization group method. The model contains ferromagnetic 
nearest-neighbor interaction $J_1^{~}$ and the competing antiferromagnetic 
interaction $J_2^{~}$. A mean-field like second-order phase transition is observed 
when the ratio $\kappa = J_2^{~} / J_1^{~}$ is less than $0.203$. In the
region $0.203 < \kappa <{1}/{4}$, the spontaneous magnetization is discontinuous
at the transition temperature. Such tricritical behavior suggests that the phase 
transitions on hyperbolic lattices need not always be mean-field like.
\end{abstract}


\maketitle

\section{Introduction}

The Ising model on the Cayley tree is known by its singular property,
where the magnetic susceptibility of the spin at the root of the tree
diverges at a temperature $T_{\rm c}^{~}$ despite the fact there is no 
singularity in the partition function of the whole system~\cite{Baxter}. 
This is a kind of phase transition which can be explained by the Ising
model on the Bethe lattice. 
It has been known that the Ising model on hyperbolic lattices, which are
negatively curved in the two-dimensional (2D) space~\cite{Sausset}, exhibits
similar aspects in common~\cite{dAuriac,Rietman,Doyon}. The universality
class of the ferromagnetic-paramagnetic phase transition of this model has been
so far considered to be mean-field like. Recent numerical studies have supported
this conjecture~\cite{Shima,Hasegawa,Ueda,Roman}.

In this paper we study effects of the antiferromagnetic next-nearest-neighbor (NNN) 
interaction $J_2^{~}$, which competes with the ferromagnetic nearest-neighbor (NN) 
one $J_1^{~}$, on the ferromagnetic-paramagnetic phase transition of the
2D Ising model on a hyperbolic lattice. We use the corner
transfer matrix renormalization group (CTMRG)
method~\cite{Nishino1,Nishino2,Nishino3}, which is a variant of the density
matrix renormalization group (DMRG) method~\cite{White1,White2,Peschel,Sch},
for the calculations of thermodynamic functions. As we show in the following, the
transition temperature $T_{0}^{~}$ monotonously decreases with the frustration
parameter $\kappa = J_2^{~} / J_1^{~}$ in the region $0 \le \kappa < {1}/{4}$,
where the ground state spin configuration is completely ferromagnetic. 
We find that there is a tricritical point when the
parameter $\kappa$ is equal to $\kappa_{\rm c}^{~} = 0.203$. 
The ferromagnetic-paramagnetic phase
transition is of the second order for $0 \le \kappa \le \kappa_{\rm c}^{~}$,
whereas it turns into the first order one for $\kappa_{\rm c}^{~} < \kappa
< {1}/{4}$.

In the next section, we explain the so-called $( 5, 4 )$ lattice in the 2D
hyperbolic space and introduce the Ising Hamiltonian on it. As a theoretical
ideal, we consider phase transition on the Bethe lattice with the coordination
number four, which is equivalent to the $( \infty, 4 )$ hyperbolic lattice.
In Sec.~III we present numerical results. Temperature dependence
of free energy, spontaneous and induced magnetizations are shown. 
We analyze these thermodynamic functions 
around the transition temperature $T_{0}^{~}$ for several values of $\kappa$, 
and determine the critical exponents $\alpha$, $\beta$, and $\delta$. 
We summarize the observed phase transition in the last section.

\section{Frustrated Ising model on Hyperbolic lattice}

\begin{figure}[tb]
\includegraphics[width=50mm,clip]{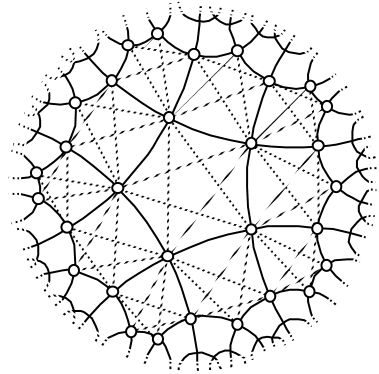}
\caption{The $( 5, 4 )$ hyperbolic lattice drawn in the Poincar\'{e} disc. The open
circles represent the Ising spins sites. The next-nearest-neighbor interactions
are here represented by the dashed lines inside the pentagons.}
\label{f1}
\end{figure}

We consider a hyperbolic 2D lattice shown in Fig.~\ref{f1},
where four pentagons share their apexes. Such lattice is conventionally called as 
the $( 5, 4 )$ lattice, where the number five represents number of the sides of each
pentagon and the number four is the coordination number.  Consider the Ising model
on this lattice, where on each lattice site labeled by $i$ there is an Ising spin
variable $\sigma_i^{~} = \pm 1$. We assume ferromagnetic interactions between NN spin
pairs shown by the full lines in Fig.~\ref{f1} and the antiferromagnetic interactions
between NNN pairs shown by the dashed lines. The Hamiltonian of the system is
represented as
\begin{equation}
{\cal H}
= - J_1^{~} \! \sum_{\langle i j \rangle = {\rm NN}}^{~} \! \sigma_i^{~} \sigma_j^{~}
+ J_2^{~} \!\!\! \sum_{\langle i k \rangle = {\rm NNN}}^{~} \!\!\! \sigma_i^{~} \sigma_k^{~}
\, ,
\label{eq1}
\end{equation}
where $J_1^{~} > 0$ is the ferromagnetic coupling constant between the NN pairs 
$\langle i j \rangle$ and $J_2^{~} > 0$ is the antiferromagnetic one between the
NNN pairs $\langle i k \rangle$. Let us define a parameter
$\kappa = J_2^{~} / J_1^{~}$ that represents strength of the frustration. 

\begin{figure}[tb]
\includegraphics[width=0.95\columnwidth,clip]{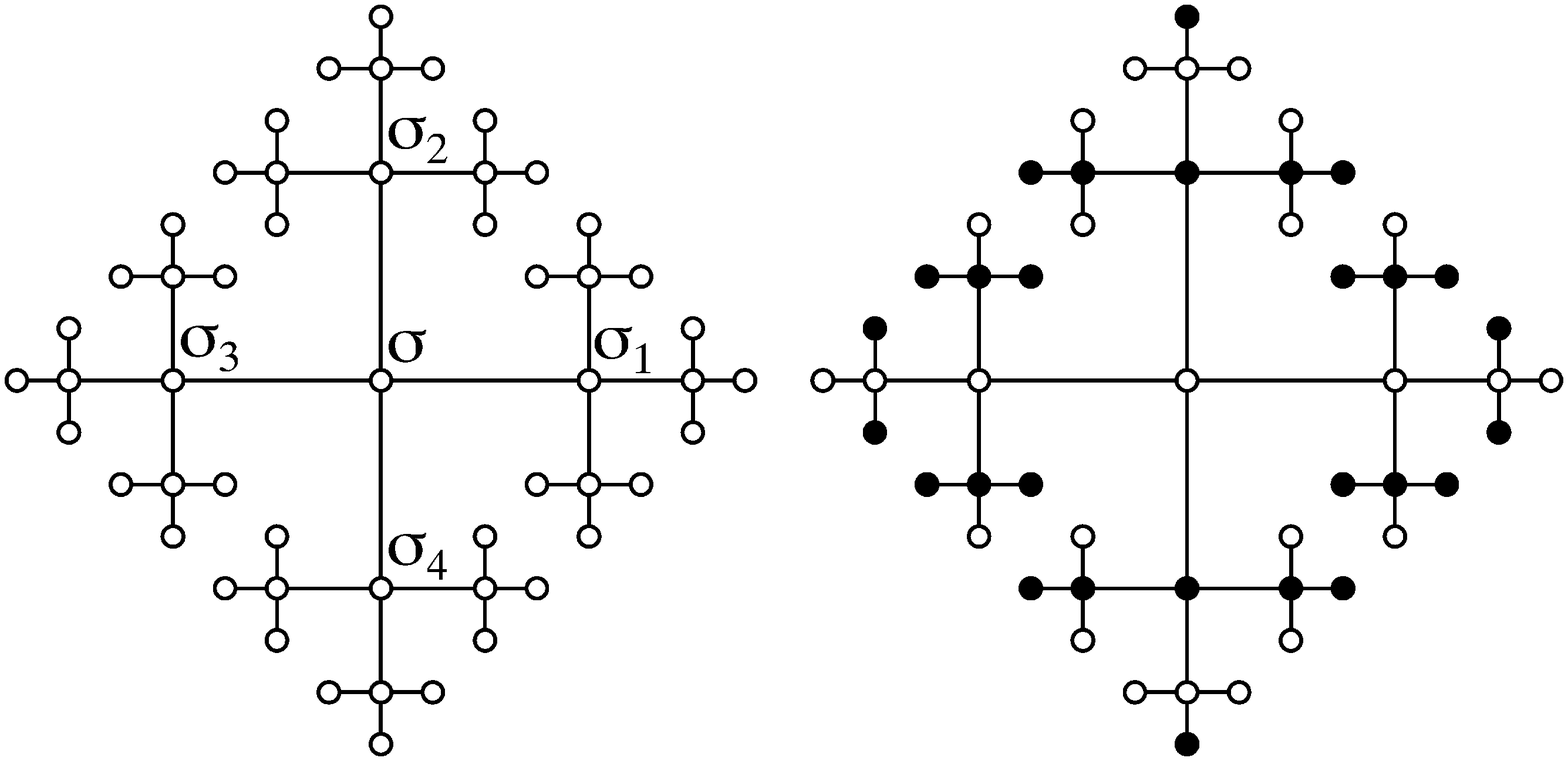}
\caption{Ground state spin configurations for $\kappa < {1}/{4}$ (left)
and $\kappa > {1}/{4}$ (right) on the $( \infty, 4 )$ lattice, which
coincides with the Bethe lattice with the coordination number four.
Note that only a finite number of spins is here depicted from the
$( \infty, 4 )$ lattice. Open circles represent spin variable $\sigma=+1$
whereas the full circles correspond to $\sigma=-1$.}
\label{f2}
\end{figure}

For purpose of obtaining brief insight of the phase structure of the $( 5, 4 )$ Ising model,
we observe the model from a wider framework. Let us introduce the
Ising model on the $( n, 4 )$ lattice where four $n$-gons (polygons
of the $n^{\rm th}$ order) meet at each lattice point. When the number
of sides $n$ ($\ge 5$) is multiple of four including the case $n = \infty$, 
the ground-state spin configuration at zero temperature is easily
obtained. Figure 2 shows the ground-state configurations for the case 
$n = \infty$, where the lattice is nothing but the Bethe lattice with
the coordination number four. For the complete ferromagnetic configuration
shown on the left, the energy expectation value per site is
\begin{equation}
\varepsilon_{\rm Ferro}^{~} = - 2 J_1^{~} + 4 J_2^{~} \, ,
\end{equation}
and for the `up-up-down-down' structure shown in the right, the value is
\begin{equation}
\varepsilon_{\rm uudd}^{~} = - 4 J_2^{~} \, .
\end{equation}
Therefore, the energy cross over $\varepsilon_{\rm Ferro}^{~} = \varepsilon_{\rm uudd}^{~}$ 
is located at $J_1^{~} = 4 J_2^{~}$, equivalently at $\kappa = {1}/{4}$. This ground state alternation
is common for all the cases where $n$ ($\ge 5$) is multiple of four. If not, the ground state 
spin configuration for large $\kappa$ is not unique and is probably disordered. 
In the case of the $( 5, 4 )$-lattice, one of the ground states in the large $\kappa$ region
can be constructed by joining the pentagons with either `up-up-up-down-down' or 
`up-up-down-down-down' spin configurations. 
After a short algebra, one obtains the energy per site
\begin{equation}
\varepsilon_{\rm uuudd}^{~} =
\varepsilon_{\rm uuddd}^{~} =
 - \frac{2}{5} J_1^{~} - \frac{12}{5} J_2^{~} 
\end{equation}
for the assumed configurations. Hence  the
energy cross over $\varepsilon_{\rm Ferro}^{~} 
= \varepsilon_{\rm uuudd}^{~}$ also occurs at $J_1^{~} = 4 J_2^{~}$.

At finite temperature, the ferromagnetic-paramagnetic phase transition is
observed in the small $\kappa$ region~\cite{AF}. Consider a single-site mean-field
approximation on arbitrary $( n, 4 )$ lattice. A mean field variable $h$ is
expressed as
\begin{equation}
h 
= ( - 4 J_1^{~} + 8 J_2^{~} ) \, \langle \sigma \rangle 
= - ( 4 - 8 \kappa ) J_1^{~} \, \langle \sigma \rangle \, ,
\end{equation}
where $\langle \sigma \rangle$ is the expectation value of the Ising spin. A
self-consistent condition for $\langle \sigma \rangle$ leads to the 
ferro\-magnetic-paramagnetic phase transition with the critical temperature 
$T_{\rm c}^{\rm M.F.}( \kappa ) = ( 4 - 8 \kappa ) J_1^{~} / k_{\rm B}^{~}$,
where $k_{\rm B}^{~}$ is the Boltzmann constant. Within this approximation, the
transition is always of the second order in the region $0 \le \kappa < {1}/{4}$,
since the effect of $J_2^{~}$ appears as the rescaling of the mean-field $h$ as in Eq.~(5).
It should be noted that $T_{\rm c}^{\rm M.F.}( \kappa = {1}/{4} ) = 2 J_1^{~} / k_{\rm B}^{~}$
is larger than zero. The mean-field approximation predicts another ordered state in the
region $\kappa > {1}/{4}$, where  the `up-up-down-down' spin configuration is 
favored if the lattice geometry allows the ordering.

An improvement to the mean-field approximation is achieved by increasing the number
of sites that are not averaged. The simplest case is the Bethe approximation, which
treats additional spins $\sigma_1^{~}$, $\sigma_2^{~}$, $\sigma_3^{~}$,
and $\sigma_4^{~}$ that surround the central site $\sigma$, as shown in Fig.~2(left). On
arbitrary $( n, 4 )$ lattice, the mean field for the surrounding four spins 
$\sigma_1^{~}$, $\sigma_2^{~}$, $\sigma_3^{~}$, and $\sigma_4^{~}$ is given by
\begin{equation}
h_a^{~} = ( - 3 J_1^{~} + 6 J_2^{~} ) \, \langle \sigma \rangle \, .
\end{equation}
As an effect of the next-nearest-neighbor interaction, the central spin $\sigma$
also feels the mean field, however, of different strength,
\begin{equation}
h_b^{~} = 8 J_2^{~} \, \langle \sigma \rangle \, ,
\end{equation}
in addition to the direct ferromagnetic interaction with the surrounding spins
$- J_1^{~} \, \sigma\, ( \sigma_1^{~} + \sigma_2^{~} + \sigma_3^{~} + \sigma_4^{~} )$.
Considering these interaction terms, one obtains the self-consistent relation
\begin{eqnarray}
\langle \sigma \rangle =  \frac{1}{Z} 
&\sum& \sigma \, 
\exp\bigl[ - \beta h_a^{~} ( \sigma_1^{~} + \sigma_2^{~} + \sigma_3^{~} + \sigma_4^{~} )
- \beta h_b^{~} \, \sigma 
\nonumber\\
&& 
+ \beta J_1^{~} \, ( \sigma_1^{~} + \sigma_2^{~} + \sigma_3^{~} + \sigma_4^{~} ) \, \sigma
\nonumber\\
&& - \beta J_2^{~} \, ( \sigma_1^{~} \sigma_2^{~} + \sigma_2^{~} \sigma_3^{~}
+ \sigma_3^{~} \sigma_4^{~} + \sigma_4^{~} \sigma_1^{~}  ) \bigr] \, ,
\label{mf1a}
\end{eqnarray}
where $\beta = 1 / k_{\rm B}^{~} T$ and where $Z$ is the partition function
\begin{eqnarray}
Z =  &\sum&
\exp\bigl[ - \beta h_a^{~} ( \sigma_1^{~} + \sigma_2^{~} + \sigma_3^{~} + \sigma_4^{~} )
- \beta h_b^{~} \, \sigma 
\nonumber\\
&& 
+ \beta J_1^{~} \, ( \sigma_1^{~} + \sigma_2^{~} + \sigma_3^{~} + \sigma_4^{~} ) \, \sigma
\nonumber\\
&& - \beta J_2^{~} \, ( \sigma_1^{~} \sigma_2^{~} + \sigma_2^{~} \sigma_3^{~}
+ \sigma_3^{~} \sigma_4^{~} + \sigma_4^{~} \sigma_1^{~}  ) \bigr] \, .
\label{mf1b}
\end{eqnarray}
The configuration sums in Eqs. (\ref{mf1a}) and (\ref{mf1b}) are taken over the 
five spins $\sigma$, $\sigma_1$, $\sigma_2$, $\sigma_3$, and $\sigma_4$.
The factorization
\begin{eqnarray}
W( \sigma_i^{~}, \sigma ) &=& \exp\left[ 
- \beta h_a^{~} \sigma_i^{~}
+ \beta J_1^{~} \sigma_i^{~} \sigma
- \beta \frac{h_b^{~}}{4} \sigma
\right] 
\end{eqnarray}
for $i = 1$, $2$, $3$, and $4$ further simplifies the expression so that the partition
function has the form
\begin{eqnarray}
Z &=&  \sum
W( \sigma_1^{~}, \sigma ) \, W( \sigma_2^{~}, \sigma ) \, 
W( \sigma_3^{~}, \sigma ) \, W( \sigma_4^{~}, \sigma )
\nonumber\\
&& \exp\bigl[ - \beta J_2^{~} \, ( 
\sigma_1^{~} \sigma_2^{~} + 
\sigma_2^{~} \sigma_3^{~} + 
\sigma_3^{~} \sigma_4^{~} + 
\sigma_4^{~} \sigma_1^{~}  ) \, \bigr] \, .
\end{eqnarray}
Since it is not a trivial task to find out an analytic solution of the
self-consistent Eqs.~(6)-(9), we solved them numerically. 
We use the parametrization $J_1^{~} = 1$ and $k_{\rm B}^{~} = 1$ 
throughout this article in the numerical calculations. Figure 3 shows
the calculated spontaneous magnetization $M = \langle \sigma \rangle$. 
The second-order phase transition is detected in the whole region
$0 \le \kappa < {1}/{4}$. As observed in the single-site mean-field 
approximation, the transition is of the second order, and the 
transition temperature remains finite even at $\kappa = {1}/{4}$.
Further improvement of the Bethe approximation can be achieved by 
means of gradual increase of unaveraged spin sites. A series
of such approximations is known as the coherent anomaly method (CAM)~\cite{CAM}. 
Here, we do not proceed with the CAM analysis; we perform
extensive numerical calculations by the CTMRG method instead.
\begin{figure}[tb]
\includegraphics[width=0.95\columnwidth,clip]{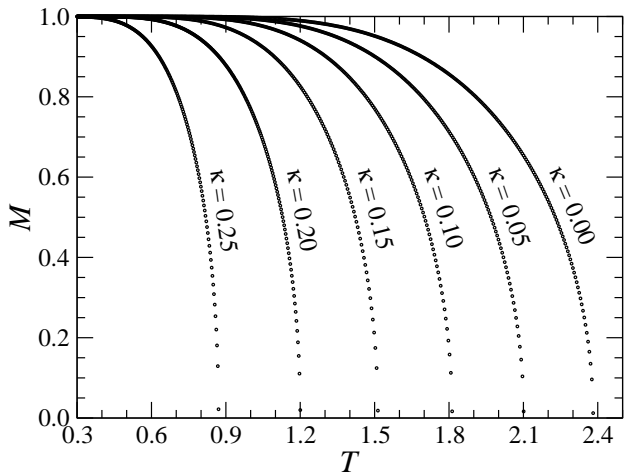}
\caption{Spontaneous magnetization $\langle \sigma \rangle$ obtained by the
Bethe approximation.}
\label{f3}
\end{figure}

It has been known that the spin expectation value $\langle \sigma \rangle$ can 
be calculated exactly at the root of the Cayley tree, which can be treated as the Bethe 
lattices~\cite{Baxter}. For the frustrated Ising model on the $( \infty, 4 )$ lattice 
shown in Fig.~2, the expectation value is expressed as
\begin{eqnarray}
\langle \sigma \rangle &=& \frac{1}{Z'} \sum \sigma \, 
W'( \sigma_1^{~}, \sigma )
W'( \sigma_2^{~}, \sigma )
W'( \sigma_3^{~}, \sigma )
W'( \sigma_4^{~}, \sigma ) \nonumber\\
&&
\exp\bigl[ - \beta J_2^{~} ( 
\sigma_1^{~} \sigma_2^{~} + 
\sigma_2^{~} \sigma_3^{~} + 
\sigma_3^{~} \sigma_4^{~} + 
\sigma_4^{~} \sigma_1^{~}  ) \bigr] 
\end{eqnarray}
with the definition of the effective partition function
\begin{eqnarray}
Z' &=& \sum 
W'( \sigma_1^{~}, \sigma )
W'( \sigma_2^{~}, \sigma )
W'( \sigma_3^{~}, \sigma )
W'( \sigma_4^{~}, \sigma ) \nonumber\\
&&
\exp\bigl[ - \beta J_2^{~} ( 
\sigma_1^{~} \sigma_2^{~} + 
\sigma_2^{~} \sigma_3^{~} + 
\sigma_3^{~} \sigma_4^{~} + 
\sigma_4^{~} \sigma_1^{~}  ) \bigr] \, ,
\end{eqnarray}
where the new factor $W'( \sigma_i^{~}, \sigma )$ 
represents a Boltzmann weight for a branch which connects
the root spin $\sigma$ with the nearest spin site $\sigma_i$.
(c.f.~Fig.~2.) This new factor $W'( \sigma_i^{~}, \sigma )$ can
be calculated from $W( \sigma_i^{~}, \sigma )$ in Eq.~(10)
repeating the application of the recursive transformation
\begin{eqnarray}
&& W_{\rm new}^{~}( \sigma_i^{~}, \sigma ) = 
\sum_{s_1^{~}, s_2^{~}, s_3^{~}}^{~}
W( s_1^{~}, \sigma )
W( s_2^{~}, \sigma )
W( s_3^{~}, \sigma ) \nonumber\\
&&
\exp\bigl[ \beta J_1^{~} \, \sigma_i^{~} \sigma - \beta J_2^{~} \, ( 
\sigma s_1^{~} + s_1^{~} s_2^{~} +  s_2^{~} s_3^{~} + s_3^{~} \sigma  ) \, \bigr] 
\nonumber\\
\end{eqnarray}
for many times until it converges~\cite{Baxter}. Thus $W'( \sigma_i^{~}, \sigma )$
contains the effect of distant sites on the Bethe lattice. Figure 4 shows
the spontaneous magnetization $M = \langle \sigma
\rangle$ calculated by Eq.~(12) using $W'( \sigma_i^{~}, \sigma )$ 
numerically obtained from Eq.~(14).  The transition is of the second order
in the region $0 \le \kappa \le 0.183$ and is of the first order in $0.184 \le \kappa
< {1}/{4}$. One can carry out perturbative calculations to ensure
that the transition temperature on the Bethe lattice is zero at
$\kappa = {1}/{4}$. The difference between Fig.~(3) and Fig.~(4) 
comes from the effect of distant interacting spin sites, which might be
essential in the tricritical behavior on the $( 5, 4 )$ lattice as studied
in the next section.

\begin{figure}[tb]
\includegraphics[width=0.95\columnwidth,clip]{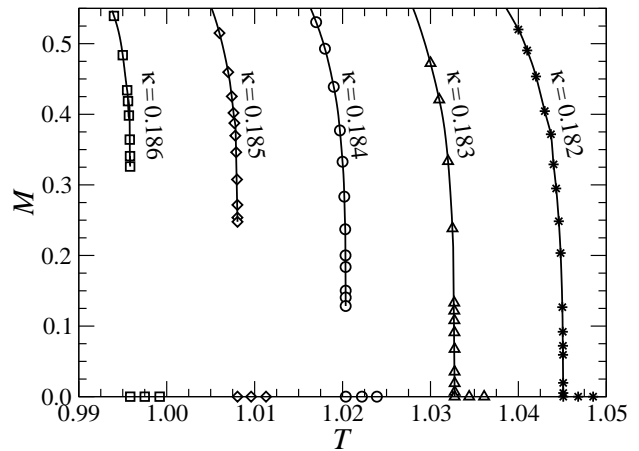}
\caption{Spontaneous magnetization of the $J_1^{~}$-$J_2^{~}$ Ising model on the
$( \infty, 4 )$ lattice. }
\label{f4}
\end{figure}

\section{Numerical results by CTMRG}

In this section we analyze the thermodynamic property of the Ising model 
on the $( 5, 4 )$ lattice by use of the CTMRG method~\cite{Nishino1,Nishino2,Nishino3}.
The method is a variant of the DMRG method~\cite{White1,White2,Sch} applied to
2D classical models~\cite{TM}. It has been known that the partition function $Z$
of a square-shaped finite-size system can be calculated as a trace of the fourth
power of the so-called corner transfer matrix (CTM), which represents a Boltzmann 
weight of a quadrant of the whole system~\cite{Baxter}. Although the matrix 
dimension of the CTM, which is denoted by $C$, increases exponentially with the linear size
of the system, it is possible to transform it into a renormalized one ${\tilde C}$
with a smaller matrix dimension $m$~\cite{kept}
by means of the RG transformation obtained from the diagonalization
of $\rho = C^4_{~}$ or $C$~\cite{Baxter,Nishino1,Nishino3}. 
This transformation is not exact but
is highly accurate in such sense that ${\tilde Z} = {\rm Tr} \, {\tilde C}^4_{~}$
is a good approximation of $Z = {\rm Tr} \, C^4_{~}$. One can precisely calculate
thermodynamic (or one-point) functions, such as the free energy $F = 
- k_{\rm B}^{~} T \log {\tilde Z}$ and the spontaneous magnetization $M$, for a
sufficiently large finite-size system by use of the CTMRG method.  Since the $( n, 4 )$
lattice can be divided into four equivalent parts (the quadrants), which share
the central site $\sigma$ on their edges, it is also possible to apply the CTMRG
method to statistical models on these lattices~\cite{Ueda,Roman,Clock}. 

In order to study critical phenomena correctly on hyperbolic lattices, we
put the following remarks. We always consider a latices system
whose linear size $L$ is several times larger than the corresponding
correlation length $\xi$, so that the central site $\sigma$ is sufficiently 
away from the system boundary. The lattice sites in the area within
the distance of the order of $\xi$ from the system boundary are affected
by the imposed ferromagnetic boundary condition, where all the Ising
spins at the system boundary point to the same direction. It should
be also noted that portion of such sites that are `near the boundary' in
the hyperbolic geometry remains finite even in the thermodynamic 
limit $L \rightarrow \infty$~\cite{Chris1,Chris2}, where the situation is similar 
to the case of the Cayley tree~\cite{Baxter}. Disregarding all these sites
`near the boundary', we focus on the thermodynamic properties of the Ising spins
deep inside the system~\cite{dAuriac,Clock}. 

\begin{figure}[tb]
\includegraphics[width=0.95\columnwidth,clip]{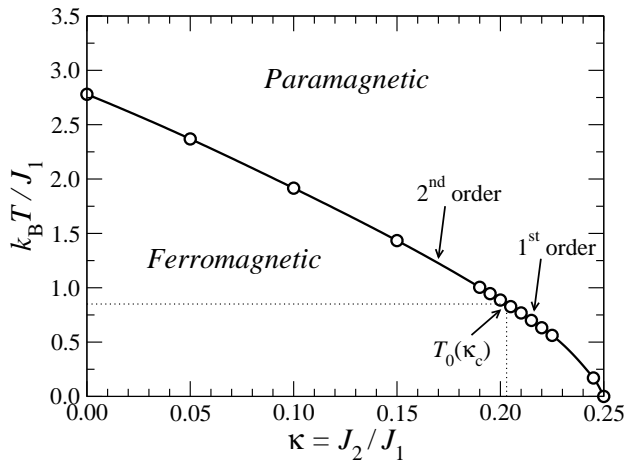}
\caption{Phase diagram of the $J_1^{~}$-$J_2^{~}$ Ising model on the 
$( 5, 4 )$ hyperbolic lattice. }
\label{f5}
\end{figure}
\begin{figure}[tb]
\includegraphics[width=0.95\columnwidth,clip]{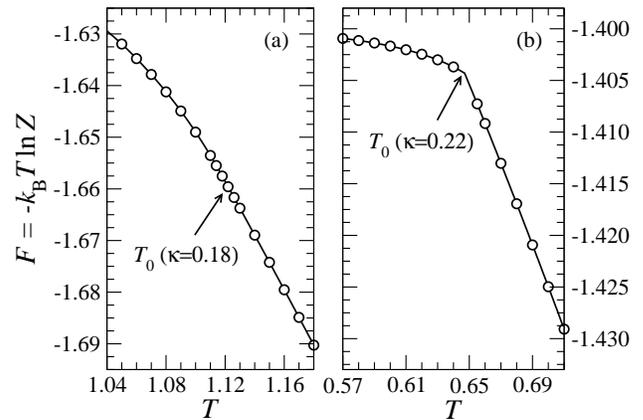}
\caption{
Dependence of the free energy $F( \kappa; T )$ on temperature when (a)
$\kappa = 0.18$ and (b) $\kappa = 0.22$. }
\label{f6}
\end{figure}

Figure~\ref{f5} shows the phase diagram
of the system in the parameter region $0 \le \kappa  < {1}/{4}$. The 
ferromagnetic-paramagnetic phase boundary is determined from the temperature 
dependence of the free energy $F( \kappa; T )$ and the spontaneous 
magnetization $M( \kappa; T )$ which we show in the following.
As an effect of the competing interactions, the transition temperature
$T_{0}^{~}( \kappa )$ monotonously decreases with $\kappa$ towards
$T_{0}^{~}( {1}/{4} ) = 0$. In the region $0 \le \kappa \le 0.203$ the transition is 
of the second order. In contrast, when $0.203 < \kappa < {1}/{4}$, we observe a 
first-order transition; the tricritical point is located at $\kappa_{\rm c}^{~} = 0.203$. 
Figure~\ref{f6} shows the
free energy $F( \kappa; T )$ at $\kappa = 0.18$ and $\kappa = 0.22$.
In the region $0.203 < \kappa < {1}/{4}$, the free energy $F( \kappa; T )$
is not a differentiable function at the transition temperature, 
as shown in~Fig.~6(b).

\begin{figure}[tb]
\includegraphics[width=0.95\columnwidth,clip]{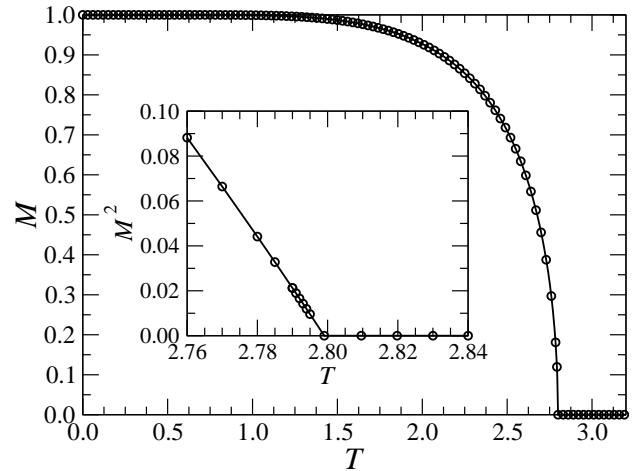}
\caption{Spontaneous magnetization $M$ for $\kappa = 0$. }
\label{f7}
\end{figure}
\begin{figure}[tb]
\includegraphics[width=0.95\columnwidth,clip]{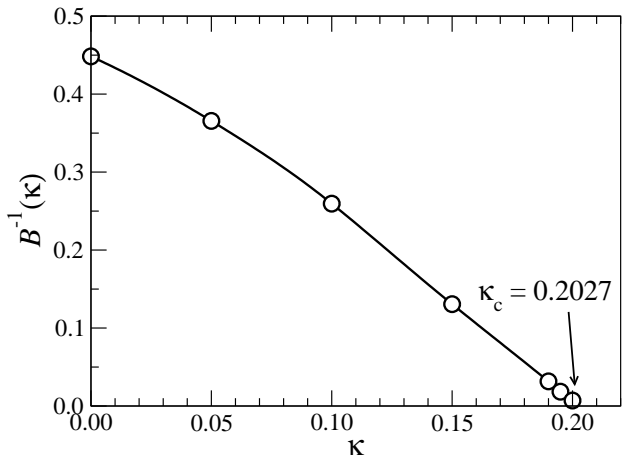}
\caption{Inverse of the prefactor $B(\kappa)$, which characterizes
the mean-field like transition observed in the spontaneous magnetization.}
\label{f8}
\end{figure}
\begin{figure}[tbp]
\includegraphics[width=0.95\columnwidth,clip]{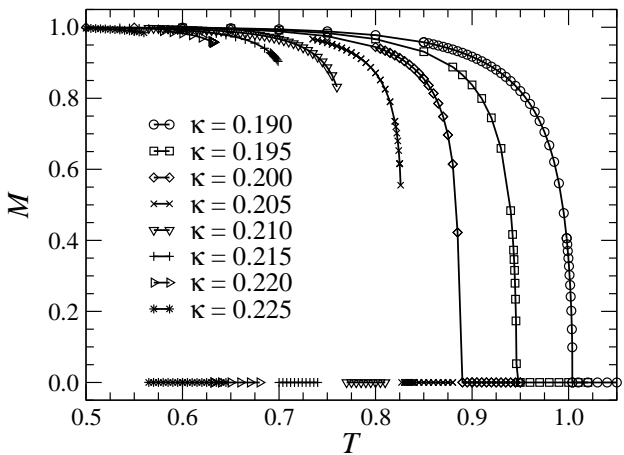}
\caption{Temperature dependence of the spontaneous magnetization
$M( \kappa; T )$ for several values $\kappa$ around $\kappa_{\rm c}^{~}$. }
\label{f9}
\end{figure}

Figure~\ref{f7} shows the spontaneous magnetization $M$ and its
square, $M^2$, when there is no frustration $\kappa = 0$. 
The squared magnetization $M^2_{~}$ is proportional
to $T_{\rm 0}^{~} - T$, the behavior which agrees with the critical
exponent $\beta = {1}/{2}$. In the second-order transition region 
$0 \le \kappa < 0.203$, the universality class remains mean-field like and
the magnetization curve satisfies the scaling form
\begin{equation}
M( \kappa; T ) = B( \kappa ) \,\left[ T_0^{~}( \kappa ) - T \right]^{1/2}_{~}
\end{equation}
around the transition temperature $T_0^{~}( \kappa )$. The prefactor $B( \kappa )$
is an increasing function of $\kappa$ and diverges at certain 
point $\kappa = \kappa_{\rm c}^{~}$. Figure~\ref{f8} shows the inverse
of the prefactor $B( \kappa )$ which linearly decreases to zero in the
vicinity of $\kappa_{\rm c}^{~}$.  We obtain $\kappa_{\rm c}^{~} = 0.2027$
from the linear fitting.

\begin{figure}[tb]
\includegraphics[width=0.95\columnwidth,clip]{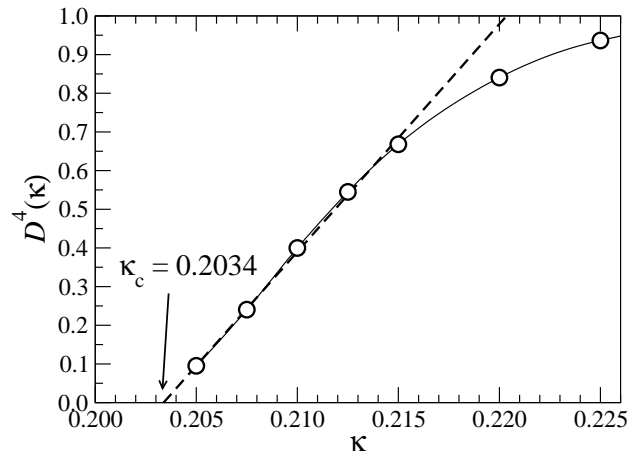}
\caption{The discontinuity function of the spontaneous magnetization at the transition 
temperature $T_0^{~}$ in the first-order transition 
region $\kappa_{\rm c}^{~} < \kappa < {1}/{4}$.}
\label{f10}
\end{figure}

One can also estimate $\kappa_{\rm c}^{~}$ out of the discontinuity
in the spontaneous magnetization $M( \kappa; T )$ in the region $0.203 <
\kappa < {1}/{4}$. Figure~\ref{f9} shows $M( \kappa; T )$ around
$\kappa = \kappa_{\rm c}^{~}$. We calculate the
discontinuity function (or the jump in the magnetization)
$D( \kappa ) = M( \kappa; T_0^{-} )$, where $T_0^{-}$ corresponds
to temperature just below the transition temperature $T_0^{~}( \kappa )$.
As shown in Fig.~\ref{f10}, the discontinuity function satisfies relation
\begin{equation}
D( \kappa ) \propto ( \kappa - \kappa_{\rm c}^{~} )^{1/4}_{~}
\end{equation}
around $\kappa = \kappa_{\rm c}^{~}$. Performing a linear fitting shown by
the dashed lines, we obtain $\kappa_{\rm c}^{~} = 0.2033$.
Comparing this value with $\kappa_{\rm c}^{~} = 0.2027$ obtained from
the data in second order region, we conclude that
the tricritical point is located at $\kappa_{\rm c}^{~} = 0.203 \pm 0.001$.

The observed tricritical behavior around $\kappa = \kappa_{\rm c}^{~}$ is 
in accordance with the Landau free energy
\begin{equation}
F( M, t ) = a \, M^6_{~} + b \, ( \kappa_{\rm c}^{~} - \kappa ) M^4_{~} + c t \, M^2 \, ,
\end{equation}
where $a$, $b$, and $c$ are positive constants or slowly varying functions of temperature.
In the second order transition region $\kappa < \kappa_{\rm c}^{~}$,
the second and the third terms in $F( M, t )$ are dominant in the vicinity of
the phase transition, and the parameter $t$ coincides with $[ T - T_0^{~}( \kappa ) ] / 
T_0^{~}( \kappa )$. 
Neglecting the first term in $F( M, t )$ below $T < T_0^{~}( \kappa )$, we 
can obtain the spontaneous magnetization $M$ that minimizes $F( M, t )$ 
from the equation $4 b \, ( \kappa_{\rm c}^{~} - \kappa ) M^2_{~} + 2 c t = 0$. 
The behavior $M^2_{~} \propto | t |$ coincides with the numerical 
result shown in Fig.~7. In the first order transition region $\kappa > \kappa_{\rm c}^{~}$,
all three terms in $F( M, t )$ are important for the minimum of the free energy. 
After short calculations, one can confirm that the jump of the spontaneous
magnetization at the transition temperature coincides with Eq.~(16), which we 
have verified from the numerical data shown in Fig.~10.

\begin{figure}[!tbp]
\includegraphics[width=0.95\columnwidth,clip]{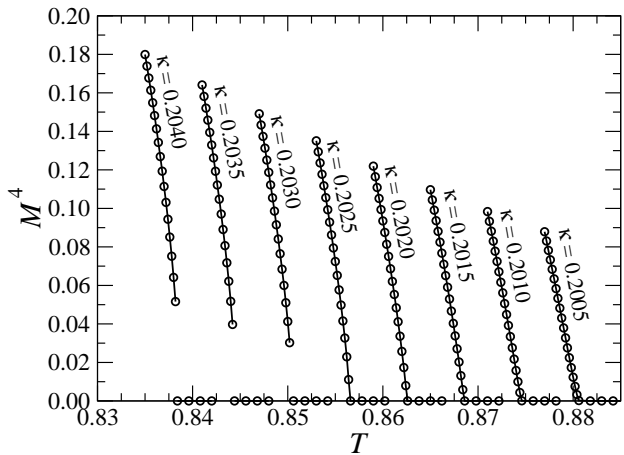}
\caption{The fourth power of the spontaneous magnetization 
around $\kappa = \kappa_{\rm c}^{~}$.}
\label{f11}
\end{figure}

\begin{figure}[tb]
\includegraphics[width=0.95\columnwidth,clip]{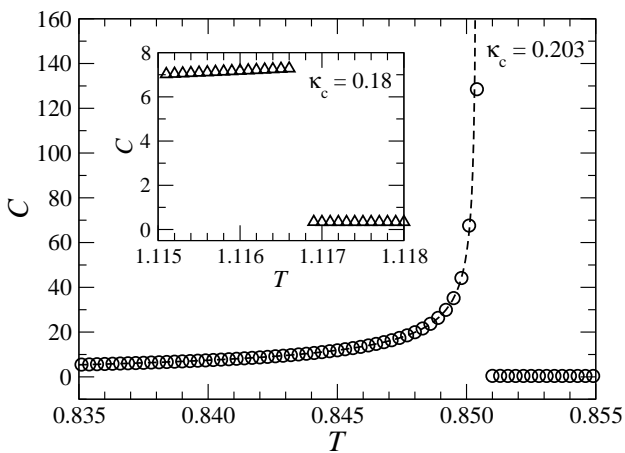}
\caption{Specific heat around the tricritical point. }
\label{f12}
\end{figure}

At the tricritical point $\kappa = \kappa_{\rm c}^{~}$, the second term in $F( M, t )$ 
in Eq.~(17) 
vanishes. Thus, the spontaneous magnetization is determined from
$6 a M^4_{~} + 2 c t = 0$, and $M^4_{~}$ is proportional to 
$T - T_{\rm 0}^{~}$. The dependence of $M^4_{~}$
with respect to temperature $T$ obtained from the numerical calculation
is shown in Fig.~\ref{f11}. The data are in accordance with
$M \propto ( T - T_{\rm 0}^{~} )^{1/4}_{~}$, which corresponds to the
exponent $\beta=1/4$ at tricriticality.  In the same 
manner it is expected that the specific heat diverges as $( T_{\rm 0}^{~} - T )^{-1/2}_{~}$,
which corresponds to the critical exponent $\alpha=1/2$. 
Figure \ref{f12} shows the numerically calculated specific heat at $\kappa_{\rm c}^{~} = 0.203$,
which agrees with the expected temperature dependence.
For comparison, in the inset of Fig.~\ref{f12}, we show the data at $\kappa = 0.18$,
which agrees with $\alpha=0$.


\begin{figure}[tb]
\includegraphics[width=0.95\columnwidth,clip]{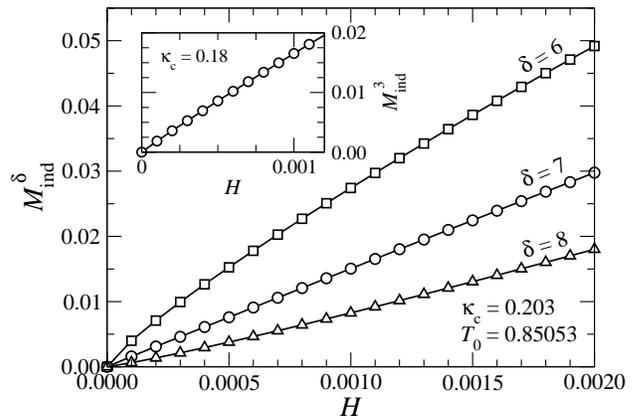}
\caption{Induced magnetization (i) at the tricritical point $\kappa = \kappa_{\rm c}^{~}$ 
and (ii) inset: that at the transition temperature when $\kappa < \kappa_{\rm c}^{~}$.}
\label{f13}
\end{figure}

Effect of an external magnetic field $H$ may be included in the Landau free energy 
by adding the interaction term $d H M$ to $F( M, t )$ in Eq.~(17), where
$d$ is a positive constant. From the assumed form of the free energy 
it is expected that $M^3_{~}$ and $M^5_{~}$ are, respectively,
proportional to $T_0^{~} - T$ when $\kappa < \kappa_{\rm c}^{~}$ and
when $\kappa = \kappa_{\rm c}^{~}$. For the confirmation, we observe the 
induced magnetization at criticality when $\kappa \leq \kappa_{\rm c}^{~}$. 
Figure~\ref{f13} shows the 
induced $M$ with respect to $H$. In the region of the second order 
phase transition, we obtained the magnetic exponent $\delta = 3$ as expected.
However, the value of the exponent $\delta$ is around $7$ at the tricritical
point, not $5$ as expected from the Landau free energy $F( M, t ) + d H M$.
Such pathological behavior in the induced magnetization is
a remaining piece of the puzzle of the current study on the $( 5, 4 )$-hyperbolic lattice.
Speak about the $J_1^{~}-J_2^{~}$ Ising model on the $( \infty, 4 )$-Bethe 
lattice, we obtain $\delta \sim 6$ at the tricritical point from 
numerical calculations. Future studies on $( n, 4 )$-lattices for $n \ge 6$ would
provide a hint to these unexpected values of $\delta$.

\section{Conclusions}

We have studied a ferromagnetic-paramagnetic phase transition of 
the $J_1^{~}$-$J_2^{~}$ Ising model on the $( 5, 4 )$ hyperbolic lattice.
A tricritical point has been found when the ratio $\kappa = J_2^{~} / J_1^{~}$,
which represents the strength of frustration, is equal to $0.203$. It should be
noted that the presence of the first-order transition cannot be obtained by
the single-site mean-field approximation applied to this system. This is in
contrast to the known fact that the phase transition of the nearest-neighbor
Ising, Potts, and clock models exhibits mean-field
nature~\cite{dAuriac,Shima,Hasegawa,Ueda,Roman,Clock}.

The observed second order phase transition in the parameter 
region $\kappa < \kappa_{\rm c}^{~} =  0.203$ belongs to the mean-field universality class,
which is characterized by the exponents $\alpha=0$, $\beta=1/2$, and $\delta=3$. 
At the tricritical point we observe $\alpha=1/2$, $\beta=1/4$, which are
in accordance with the Landau free energy written in the even polynomial of
order parameter. The observed value of exponent  $\delta\approx7$ at the 
tricritical point is the only exception which requires further detailed studies.

As an effect of the frustration, entropy of the ordered phase shall be enhanced
compared with the ordered state that has the same spontaneous magnetization under
$J_2^{~} = 0$. We conjecture that this enhancement effect creates a minimum in 
the Landau free energy, which may be the reason of the first order transition we
have observed here. 
The tricritical point is also present in the $( \infty, 4 )$ lattice, which is nothing
but the Bethe lattice. This suggests that the suppression of the loop back effect
in the hyperbolic lattice is essential for the appearance of the tricritical behavior.

Determination of the phase diagram in the region $\kappa > {1}/{4}$ is
challenging because the ground state spin configuration becomes
non-trivial as has been discussed in Section II. 
Because the $( 5, 4 )$ hyperbolic lattice consists 
of pentagons,
the lattice does not decouple into sub-lattices even when $J_1^{~} = 0$.
For the study of this region, we have to modify the CTMRG algorithm
in order to treat ordered states with non-trivial spin patterns.

\section*{Acknowledgments}

Slovak Agency for Science and Research grant APVV-51-003505 and
Slovak VEGA grant No. 2/6101/27 are acknowledged (A.G. and R.K.). This
work is also partially supported by Grant-in-Aid for Scientific Research from
Japanese Ministry of Education, Culture, Sports, Science and Technology
(T.N. and A.G.). A.G. acknowledges the support of the Alexander von
Humboldt foundation.


\begin{thebibliography}{99}
\bibitem{Baxter} R.~J.~Baxter, {\it Exactly solved models in statistical mechanics} 
Academic Press, London (1982).
\bibitem{Sausset} F.~Sausset  and G.~Tarjus  {\it J. Phys. A: Math. Gen.} {\bf 40}
12873 (2007).
\bibitem{dAuriac} J.C.~Angl\'es d'Auriac, R.~M\'elin, P.~Chandra  and B.~Dou\c{c}ot 
{\it J. Phys. A: Math. Gen.} {\bf 34} 675 (2001).
\bibitem{Rietman} R.~Rietman, B.~Nienhuis, and J.~Oitmaa, {\it J. Phys. A} {\bf 25}
 6577 (1992).
\bibitem{Doyon} B.~Doyon  and P.~Fonseca {\it J. Stat. Mech.} P07002  (2004).
\bibitem{Shima} H.~Shima and Y.~Sakaniwa, {\it J. Phys. A} {\bf 39} 4921(2006).
\bibitem{Hasegawa} I.~Hasegawa, Y.~Sakaniwa, and H.~Shima, 
{\it Surface Science} {\bf 601} 5232 (2007).
\bibitem{Ueda} K.~Ueda, R.~Krcmar, A.~Gendiar, and T.~Nishino, 
{\it J. Phys. Soc. Jpn.} {\bf 76} 084004 (2007).
\bibitem{Roman} R.~Krcmar, A.~Gendiar, K.~Ueda, and T.~Nishino, {\it J. Phys. A:
Math. Theor.} {\bf 41} (2008) 125001.
\bibitem{Chris1} N.~Anders and C.~Chris Wu, {\it Combinatorics, Probability 
and Computing} {\bf 14} 523 (2005).
\bibitem{Chris2} C.~Chris Wu, {\it J. Stat. Phys.} {\bf 100} 893 (2000).
\bibitem{Clock} A.~Gendiar, R.~Krcmar, K~Ueda, and T.~Nishino, {\it Phys. Rev. E} {\bf 77}
041123 (2008).
\bibitem{Nishino1} T.~Nishino and K.~Okunishi, {\it J. Phys. Soc. Jpn.} {\bf 65}
891 (1996).
\bibitem{Nishino2} T.~Nishino, K.~Okunishi, and M.~Kikuchi, {\it Phys. Lett. A} 
{\bf 213} 69 (1996).
\bibitem{Nishino3} T.~Nishino and K.~Okunishi, {\it J. Phys. Soc. Jpn.} {\bf 66}
3040 (1997).
\bibitem{White1} S.~R.~White, {\it Phys. Rev. Lett.} {\bf 69} 2863 (1992).
\bibitem{White2} S.~R.~White, {\it Phys. Rev. B} {\bf 48} 10345 (1993).
\bibitem{Sch} U.~Schollw\"{o}ck, {\it Rev. Mod. Phys.} {\bf 77} 259 (2005).
\bibitem{Peschel} I.~Peschel, X.~Wang, M.~Kaulke, and K.~Hallberg (Eds.), 
{\it Density-Matrix Renormalization, A New Numerical Method in Physics}, 
Lecture Notes in Physics (Springer, Berlin 1999).
\bibitem{AF} If $n$ is multiple of 4, ferromagnetic-antiferromagnetic transition would be 
observed in the large $\kappa$ region, which is outside the scope of this article.
\bibitem{CAM} M.~Suzuki and M.~Katori, J. Phys. Soc. Jpn. {\bf 55} 1 (1986).
\bibitem{TM} T.~Nishino, J. Phys. Soc. Jpn. {\bf 64} (1995) 3598.
\bibitem{kept} We keep $m = 40$ block spin states at most in the numerical
calculations.
\end{thebibliography}
\end{document}